\documentclass[twocolumn,showpacs,preprintnumbers,amsmath,amssymb]{revtex4}
\usepackage{graphicx}% Include figure files
\usepackage{dcolumn}% Align table columns on decimal point
\usepackage{bm}% bold math

\def\mr#1{\mathrm{#1}}

\DeclareMathOperator{\ineqr}{ }
\DeclareMathOperator{\ineql}{ }
\def\arangle{\stackrel{>}{\ineqr\limits_{\sim}}}
\def\alangle{\stackrel{<}{\ineql\limits_{\sim}}}
\makeatletter
\@addtoreset{footnote}{page}
\makeatother
 
\begin{document}

\preprint{}

\title{\normalsize{High Energy Neutrino Flashes from Far-Ultraviolet and X-ray
Flares in Gamma-Ray Bursts}}% Force line breaks with \\

\author{Kohta Murase$^{1}$}
%\email{kmurase@yukawa.kyoto-u.ac.jp}%
\author{Shigehiro Nagataki$^{1,2}$}%
%\email{nagataki@yukawa.kyoto-u.ac.jp}
\affiliation{%
$^{1}$Yukawa Institute for Theoretical Physics, Kyoto University,
Oiwake-cho, Kitashirakawa, Sakyo-ku, Kyoto, 606-8502, Japan \\
$^{2}$KIPAC, Stanford University, P.O.Box 20450, MS 29,
Stanford, CA, 94309, USA
}%

%\date{\today}% It is always \today, today,
           %  but any date may be explicitly specified
                        
\begin{abstract}
\small{
The recent observations of bright optical and x-ray flares by the
Swift satellite suggest these are
produced by the late activities of the central engine. We study the
neutrino emission from far-ultraviolet and x-ray flares under the late 
internal shock model. We show that the efficiency of pion production
in the highest energy is
comparable to or higher than the unity, and the contribution from such 
neutrino flashes to a diffuse very high energy neutrino background can
be larger than that of prompt bursts if the total baryonic energy input
into flares is comparable to the radiated energy of prompt bursts. 
These signals may be detected by IceCube and are very important
because they have possibilities to probe the nature of flares (the
 baryon loading, the photon field, the magnetic field and so on).} 
\end{abstract}

% PACS, the Physics and Astronomy
                              % Classification Scheme.
                              %\keywords{Suggested keywords}%Use showkeys class option if keyword		
                              %display desired
\maketitle

\normalsize{
High energy neutrino emission from gamma-ray bursts (GRBs) has been
expected in the context of the standard internal-external scenario of
GRBs. Especially, since the prediction of Waxman and Bahcall
\cite{Wax1}, neutrino bursts in the internal shock model have
been studied by several authors and neutrino afterglows in the external shock
model have also been discussed \cite{KM1,Wax2}.

The standard model of GRBs has succeeded in explaining of many
observations, but there were a few outstanding questions in the
study of GRBs before the launch of the Swift satellite (see reviews,
e.g., \cite{Zha1}). The Swift satellite, which is an 
ideal mission to answer these questions, 
has presented indeed very fruitful results
during the first several months of its operation. Especially in the
x-ray band, Swift x-ray telescope detected an x-ray afterglow
for essentially
every burst, which showed the surprising behaviors that are not
straightforwardly expected in the pre-Swift era \cite{Zha2}. 
The early afterglow light curve has several surprising features such as
an early steep decay, a follow-up shallower-than-normal decay, one
or more x-ray flares and so on (see e.g., \cite{Zha2}). 
Especially, observations show that many bursts have large x-ray flares 
superimposed on the underlying afterglow. In the early afterglows of
XRF 050406 and GRB 050502b, x-ray telescope detected mysterious strong
x-ray flares,
and some flares such as GRB 050607 and GRB 050904 have multiple flares
\cite{Bur1}. These observational results suggest the existence of
additional emission in the early afterglow phase besides the
conventional forward shock emission.

In this Letter we calculate high energy neutrino emission from
far-ultravioet (FUV) and x-ray flares under the late internal shock
model \cite{Fan1}. Our method of
calculation using GEANT4 \cite{Ago1} is the same as in Murase \&
Nagataki \cite{KM1}, but quantitatively improved \cite{PDG1}.
Now, large neutrino detectors such as IceCube, ANTARES, and NESTOR are
being constructed \cite{Ahr1}. In the near future, 
these detectors may detect high energy neutrino signals correlated
with flares.

\textit{The model}. The flares typically happen hundreds of seconds after the 
trigger of prompt emissions or
earlier. In some cases, they occur 
around a day after the burst.
The observed typical time scale is larger than that of the prompt
emission, which is $\delta t \sim (10-{10}^{3})$ s \cite{Zha2,Bur1}.
The amplitudes of the flares are usually larger than the underlying
afterglow component by a factor of several, but can be much
larger. These can be even comparable to or higher
than  
the prompt burst component which is typically  $L_\mr{{X}}^{\mr{GRB}}
\sim ({10}^{49}-{10}^{52})$ ergs/s around ($1 - 10$) keV band. Hereafter 
we take  $L_{\mr{max}} = ({10}^{47}-{10}^{50}) \, \mr{ergs/s}$ as a peak 
luminosity of far-ultraviolet and x-ray flares. Although 
some flares allow for the possibility of external shock processes,
the general features of the flares suggest that this phenomenon is
best interpreted as  a late internal central
engine activity \cite{Zha2}. The variability of some GRB afterglows
implies that the engine may last much longer than the duration of the
bursts \cite{Iok1}, although a possible mechanism for reactivity of
the central engine is unknown \cite{Pro1}. 
We suppose the late internal shock occurs a few minutes after the
prompt $\gamma$-ray emission, powering a new unsteady relativistic
outflow. Falcone \textit{et al}. show that the case of GRB 050502b implies 
the late outflow has the smaller Lorentz factor, $\Gamma \alangle 20$ 
\cite{Bur1}. Hence, we assume that the typical
Lorentz factors of the ejected material are smaller than the prompt
emission, setting $\Gamma _{\mr{s}}\sim 10$ and $\Gamma _{\mr{f}} \sim
100$ as the typical Lorentz factors of the slow and fast shells,
respectively. We can estimate the Lorentz factor of the merged shell 
$\Gamma \approx
\sqrt{\Gamma _\mr{f} \Gamma _\mr{s}} \simeq 30$, and the Lorentz
factor of the internal shocks can be estimated by
$\Gamma _\mr{sh} \approx
(\sqrt{\Gamma_{\mr{f}}/\Gamma_{\mr{s}}}+\sqrt{\Gamma_{\mr{s}}/\Gamma_{\mr{f}}})
/2 \sim$ a few. This regenerated internal shocks are called as the
late internal shocks \cite{Fan1}. The typical collision radius is
expressed by commonly used relation, 
$r \approx {10}^{14.5}{(\Gamma/30)}^2 {[\delta t/6(1+z) \, \mr{s}]}$
cm. Of course, it should be smaller than the deceleration radius.
Actually the exact radiation mechanism producing the flares is
unclear. Here, however, let us assume that flares are
produced by the synchrotron emission to see typical parameters,. The
minimal Lorentz factor of electrons is estimated by $\gamma
_{e,m} \approx \epsilon _e (m_p/m_e)(\Gamma _{\mr{sh}}-1)$.
Since we can estimate the intensity of magnetic field by $B = 7.7
\times {10}^{3} \, \mr{G} \epsilon _{B,-1}^{1/2} {(\Gamma _{\mr{sh}}
(\Gamma _{\mr{sh}}-1)/2)}^{1/2} L _{\mr{M},50}^{1/2} \Gamma
_{30}^{-1} r_{14.5}^{-1}$, the observed break energy is,
$E ^{b}= \hbar \gamma_{e,m}^2 \Gamma eB/m_e c
\sim 0.1 \, \, \mr{keV} \epsilon _{e,-1}^2 \epsilon _{B,-1}^{1/2} 
{(\Gamma _{\mr{sh}}-1)}^{5/2} {(\Gamma _{\mr{sh}}/2)}^{1/2} 
L _{\mr{M},50}^{1/2} r_{14.5}^{-1}$ \cite{Fan1}.
Here, $L_{\mr{M}}$ is the outflow luminosity.
Therefore, the typical emitted energy is in the soft x-ray band.

For numerical calculations, we simply set the break energy and adopt the
power-law spectrum similarly to that of the case of the prompt
emission. Although most of flares are actually well fitted by a Band
function or cutoff power-law model \cite{Bur1}, such a treatment
does not change our results so much. We use the following expression
in the comoving frame, $dn/d\varepsilon = n_b{(\varepsilon/
{\varepsilon}^{b})}^{-\alpha}$ for $\varepsilon^{\mr{min}} 
< \varepsilon < \varepsilon ^{b}$ or $dn/d\varepsilon=n_b
{(\varepsilon/{\varepsilon}^{b})}^{-\beta}$ for $\varepsilon
^b < \varepsilon < \varepsilon ^{\mr{max}}$, 
where we set $\varepsilon ^{\mr{min}}=0.1$ eV because the synchrotron
self-absorption will be crucial below this energy and
$\varepsilon ^{\mr{max}}=1$ MeV because the pair absorption will be
crucial above this energy \cite{Li1}.     
Actually we do not know the peak energy and the lower spectral index
of many flares \cite{Fan1}. So we allow for the existence of
far-ultraviolet-ray (FUV-ray) flares and take
$\varepsilon ^{b}=(0.01-0.1)$ keV in the comoving frame. We assume 
$\alpha=1$ and set $\beta =2.2$ \cite{Zha2} similarly to the prompt 
emission.
The photon energy density is $U_{\gamma} =\int d\varepsilon \, 
\varepsilon dn/d\varepsilon $. The magnetic energy density and the nonthermal
proton energy density are expressed by $U_{B}= \xi_{B} U_{\gamma}$ and
$U_{p}=\xi_{\mr{acc}}U_{\gamma}$, respectively. The nonthermal baryon
loading factor $\xi_{\mr{acc}}$ can be expressed by $\xi_{\mr{acc}}
\approx \zeta _{p}10(0.1/\epsilon _{e})$, where $\zeta 
_{p}$ is the proton acceleration efficiency. Roughly speaking, 
$\xi_{\mr{acc}}=10$ and $\xi _{B}=1$ correspond to $\epsilon _{B}=0.1$ 
if $\zeta _{p} \sim 1$. We take $r \sim {10}^{14.5-16}$ cm 
with $\Gamma \sim (10-50)$. Although we have done wide parameter 
surveys, we will hereafter set the width of the shell to 
$r/2\Gamma ^2 = 4.5 \times {10}^{11-12}$ cm 
according to $\delta t =(30-300)$ s at $z=1$, and show the two cases 
of $r = {10}^{14.9}$ cm with $\Gamma=30$ and $r={10}^{15.3}$ cm with 
$\Gamma=15$. 

To obtain the pion production rate and estimate the maximal energy of
accelerated protons, we need to take into account following various cooling
time scales. We consider the synchrotron cooling time written by 
$t_\mr{syn}=3m_p^4 c^3/4 \sigma _\mr{T} m_e^2 \varepsilon _{p} U_B$,
the inverse-Compton (IC) cooling time which is given by
Jones \cite{Jon1}, the adiabatic cooling time comparable to the
dynamical time, and the photomeson cooling time which is evaluated by,  
\begin{equation}
t^{-1}_{p\gamma}(\varepsilon _{p}) = \frac{c}{2{\gamma}^{2}_{p}} \int_{\bar{\varepsilon}_{\mr{th}}}^{\infty} \! \! \! d\bar{\varepsilon} \, 
{\sigma}_{p\gamma}(\bar{\varepsilon}) {\kappa}_{p}(\bar{\varepsilon})
\bar{\varepsilon} \int_{\bar{\varepsilon}/2{\gamma}_{p}}^{\infty} \! \! \! \! \! \! \! \! \! d \varepsilon \, {\varepsilon}^{-2} 
\frac{dn}{d\varepsilon}, \label{pgamma}
\end{equation}
where $\bar{\varepsilon}$ is the photon energy in the rest frame of
proton, $\gamma _{p}$ is the proton's Lorentz factor, $\kappa _{p}$ is
the inelasticity of proton, and $\bar{\varepsilon} _{\mr{th}} \approx
145$ MeV is the
threshold photon energy for photomeson production.
From various time scales, we can estimate the total cooling time scale
by $t_{p}^{-1} \equiv t_{p\gamma}^{-1} + t_{\mr{syn}}^{-1} +
t_{\mr{IC}}^{-1} + t_{\mr{ad}}^{-1}$.
We believe that not only electrons but also protons can be accelerated
by the first-order Fermi acceleration mechanism and assume $dn_{p}/d
\varepsilon _{p} \propto \varepsilon _{p}^{-2}$. 
By the condition $t_{\mr{acc}}<t_{p}$, we can estimate the
maximal energy of accelerated protons. Here, we take the
acceleration time scale by $t_{\mr{acc}} \sim \varepsilon
_{p}/eBc$. We also set the minimal energy of protons to
$10$ GeV because this would be around $\sim \Gamma _{\mr{sh}} m_p c^2$. One
of numerical results is shown in Fig. 1.\\
\begin{figure}[b]
\includegraphics[width=0.65\linewidth]{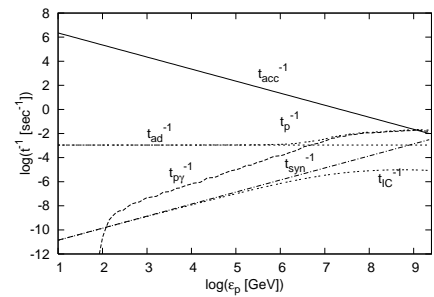}% Here is how to import EPS art
\caption{\footnotesize{\label{Fig1} Various cooling time scales and the acceleration
time scale for FUV-ray flares with $L_{\mr{max}} = {10}^{49} \,
\mr{ergs/s}$, $\xi _{B}=1$, and $r={10}^{14.9}$ cm with $\Gamma
=30$. Note that energy scale is measured
in the shell comoving frame.}}
\end{figure}
Let us evaluate $f_{p\gamma} \equiv t_{\mr{dyn}}/t_{p\gamma}$
by $\Delta$-resonance approximation. After performing the second
integral in Eq. (\ref{pgamma}), we can approximate by,
\begin{eqnarray}
t_{p\gamma}^{-1} \simeq \frac{U_{\gamma}}{2 \varepsilon ^b}
\!\!\! &c& \!\!\! \sigma _{\Delta} \, \kappa _{p}(\bar{\varepsilon}
_{\Delta}) \frac{\Delta \bar{\varepsilon}}{\bar{\varepsilon} _{\Delta}}
 \left\{ \begin{array}{rl} 
{(\bar{\varepsilon} _{\Delta}/2 \gamma _p \varepsilon
^b)}^{-(\beta-1)}\\
{(\bar{\varepsilon} _{\Delta}/2 \gamma _p \varepsilon
^{b})}^{-(\alpha-1)} 
\end{array} \right. , \label{pgamma2}
\end{eqnarray}
where $\sigma _{\Delta} \sim 5 \times {10}^{-28} \, {\mr{cm}}^2$,
$\kappa _{p}(\bar{\varepsilon} _{\Delta}) \sim 0.2$, and $\bar{\varepsilon}
_{\Delta} \sim 0.3$ GeV, and $\Delta \bar{\varepsilon} \sim 0.2$  
GeV \cite{Wax1}.
Here, we have included the effect of multi-pion production
and high inelasticity which is moderately important, and multiplied by
a factor of $\sim (2-3)$ in  Eq. (\ref{pgamma2}) \cite{KM1}. Hence, we
can obtain
\begin{equation}
f_{p\gamma} \simeq 10 \frac{L_{\mr{max},49}}{r_{14.5} \Gamma _{30}^2
E_{\mr{keV}}^{b}} \left\{ \begin{array}{rl} 
{(E_p/E_p^b)}^{\beta-1} & \mbox{($E_p < E_{p}^{b}$)}\\
{(E_{p}/E_{p}^{b})}^{\alpha-1} & \mbox{($E _p^{b} < E_p$)} 
\end{array} \right., \label{pgamma3}
\end{equation}
where $E_{p}^{b} \simeq 0.5 \bar{\varepsilon}
_{\Delta}m_pc^2 \Gamma ^2/E^{b}$ is the
proton break energy.
From Eq. (\ref{pgamma3}), 
we can conclude that a significant
fraction of high energy accelerated protons 
cannot escape from the source without photomeson productions. In the case
of bright x-ray flares whose luminosity is
larger than ${10}^{49}$ ergs/s and FUV-ray flares, 
almost all
protons accelerated to the very high energy region will be depleted.\\
\begin{figure}[t]
\includegraphics[width=0.65\linewidth]{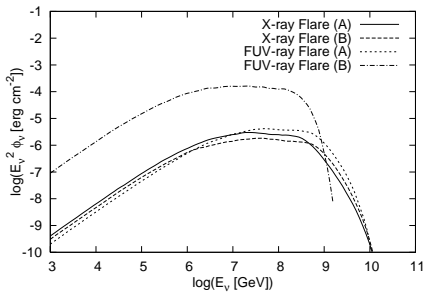}% Here is how to import EPS art
\caption{\footnotesize{\label{Fig2} The observed muon-neutrino $(\nu _{\mu} +
\bar{\nu} _{\mu})$ spectra for one GRB event at $z=0.1$. X-ray Flare (A):
$L_{\mr{max}} = {10}^{49} \, \mr{ergs/s}$, $\xi _{B}=1$,
$\xi_{\mr{acc}}=10$, and $r={10}^{14.9}$ cm with $\Gamma
=30$. X-ray Flare (B): $L_{\mr{max}} = {10}^{48} \, \mr{ergs/s}$, $\xi
_{B}=1$, $\xi_{\mr{acc}}=10$, and $r={10}^{15.3}$ cm with
$\Gamma =15$. FUV-ray Flare (A): $L_{\mr{max}} = {10}^{48} \,
\mr{ergs/s}$, $\xi _{B}=1$, $\xi_{\mr{acc}}=10$, and $r={10}^{15.3}$ cm with
$\Gamma =15$. FUV-ray Flare (B): $L_{\mr{max}} = {10}^{50} \, \mr{ergs/s}$, 
$\xi _{B}=0.1$, $\xi_{\mr{acc}}=30$, and $r={10}^{14.9}$ cm with
$\Gamma =30$.}}
\end{figure}
\textit{Neutrino spectrum and flux}. As in the case of the prompt
emission, we can
expect high energy neutrino flashes from one GRB event only if the
flare is nearby or energetic. In Fig. 2, we show an example of the observed
neutrino flux from the source at $z=0.1$. The expected muon events for
above TeV energy neutrinos are $N _{\mu}=0.02$ events in the case of
x-ray Flare (A) in Fig. 2. We can expect $N _{\mu}=1.4$ events when
the flare is energetic and more nonthermally baryonic [FUV-ray Flare
(B)]. Note that, in the case of an energetic flare, the maximal
neutrino energy will become small because the highest energy
protons suffer from the photomeson cooling very much.

We can estimate a diffuse neutrino background from FUV/x-ray flares for
specific parameter sets under the standard $\Lambda$CDM cosmology
$(\Omega _{\mr{m}}=0.3, \Omega _{\Lambda}=0.7; H_{0}=71 \, \mr{km
\, s^{-1} \, Mpc^{-1}})$. Assuming that the long GRB
rate traces the starformation rate (SFR), we shall use the SF3
model of Porciani and Madau combined with the 
normalization of overall GRB rates obtained by Guetta \textit{et al}. 
\cite{Por1}. The
choice of $z_{\mr{max}}$ which should be larger than $6.3$ does not
affect our results so much, so we take $z_{\mr{max}}=11$, which is the
epoch when the reionization is likely to occur.
X-ray flares are detected in at least (1/3 - 1/2) of Swift GRBs
\cite{Zha2}. So we can expect that flares are common in GRBs.
Here, introducing the ratio of the energy emitted by flares
to that of the prompt emission, $f_{\mr{F}} \equiv E_{\mr{flare}}/
E_{\mr{GRB}}$, let us estimate the neutrino flux analytically. 
First, we can express the total number spectrum of accelerated protons, using 
$\varepsilon _{p,\mr{max}} \sim {10}^{9}$ GeV,
\begin{equation}
E_p^2\frac{dN_p}{dE_p} \simeq 1.6 \times {10}^{50} N f_{b} \xi
_{\mr{acc},10} L_{\mr{max},49} \left(\frac{r_{14.5}}{\Gamma _{30}^2}
\right) \, \mr{erg},
\end{equation}
where $f_{b}$ is the beaming factor and $N$ is the number of flares.
Since we have already estimated $f_{p\gamma}$ in 
Eq. (\ref{pgamma3}), 
by replacing $2.5 N f_{b} L_{\mr{max}}r/\Gamma ^2 c$ with the
total emitted energy from flares $E_{\mr{flare}}$, we can estimate the diffuse
neutrino background flux from flares as follows,
\begin{eqnarray}
E_{\nu}^2 \Phi _{\nu} &\sim& \frac{c}{4\pi H_{0}}
 \frac{1}{4} \mr{min}[1,f_{p\gamma}] E_{p}^2 \frac{dN_{p}}{dE_{p}}
 R_{\mr{GRB}}(0) f_{z} \nonumber\\
&\simeq& 6 \times 10^{-10} \mr{GeV cm^{-2} s^{-1} str^{-1}}
 \mr{min}[1,f_{p\gamma}]  \nonumber\\
&\times&  f_{\mr{F}} \xi _{\mr{acc}} E _{\mr{GRB},51} 
\left(\frac{R _{\mr{GRB}}(0)}{20 \,
 \mr{Gpc}^{-3}\mr{yr}^{-1}}\right) \left( 
\frac{f_{z}}{3} \right),
\end{eqnarray}
where $f_{z}$ is the correction factor for the possible contribution from
high redshift sources and $R_{\mr{GRB}}(0)$ is the overall GRB rate at $z=0$
where the geometrically correction is taken into account. 
$f_{\mr{F}}\xi _{\mr{acc}}$ expresses the ratio of the nonthermal
baryon energy of flares to the prompt radiated energy. If the
nonthermal baryon  energy is comparable to $E_{\mr{GRB}}$, we can expect a
significant contribution to the neutrino background from flares. 
Our numerical results obtained by the same method as in our previous
paper are shown in Fig. 3 for the cases of $f _{\mr{F}} \xi
_{\mr{acc}}= (0.5-5)$. Expected muon events from above TeV neutrinos are
also shown in Fig. 3. Although the expected muon events of above TeV
neutrinos from flares will be smaller than those from prompt bursts
(shown in Fig. 3), they can be comparable with or exceed those from prompt
bursts in the very high energy region above a few PeV range.\\
\begin{figure}[b]
\includegraphics[width=0.65\linewidth]{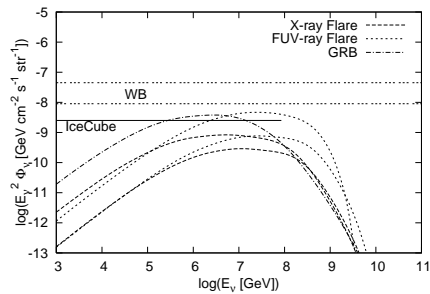}% Here is how to import EPS art
\caption{\footnotesize{\label{Fig3} The neutrino background from 
flares. X-ray Flare; (the upper dashed line): $L_{\mr{max}}={10}^{49} \, 
\mr{ergs/s}$, $\xi _{B}=1$, $f_{\mr{F}} \xi _{\mr{acc}}=1$, and
$r={10}^{15.3}$ cm with $\Gamma =15$; $N_{\mu}= 3.7$ events/yr.
(The lower dashed line): $L_{\mr{max}}={10}^{49} \,
\mr{ergs/s}$, $\xi _{B}=1$, $f_{\mr{F}} \xi_{\mr{acc}}=0.5$, and
$r={10}^{14.9}$ cm with $\Gamma =30$; $N_{\mu}=0.8$ events/yr. FUV-ray
Flare; (the upper dotted line): $L_{\mr{max}}={10}^{49} \,
\mr{ergs/s}$, $\xi _{B}=0.1$, $f_{\mr{F}} \xi _{\mr{acc}}=5$, and
$r={10}^{14.9}$ cm with $\Gamma =30$; $N_{\mu}=8.8$ events/yr.
(The lower dotted line): $L_{\mr{max}}={10}^{48} \, 
\mr{ergs/s}$, $\xi _{B}=1$, $f_{\mr{F}} \xi _{\mr{acc}}=1$, and 
$r={10}^{15.3}$ cm with $\Gamma =15$; $N_{\mu}=1.4$ events/yr. GRB: taken
from \cite{KM1} with $E_{\gamma,\mr{sh}}^{\mr{iso}}=2 \times
{10}^{51} \, \mr{ergs}$, $\xi _{B}=1$, $\xi _{\mr{acc}}=10$, and 
$r= ({10}^{13} - {10}^{14.5}) \, \mr{cm}$; $N_{\mu}=21$ events/yr. 
WB: Waxman-Bahcall bounds \cite{Wax1}.}}
\end{figure}  
\textit{Implications and discussions}. In the near future, 
high energy neutrino signals from flares may be detected by IceCube
and/or Auger which can provide information on the nature of flares. We
expect such neutrino flashes from flares should be in coincidence with
the early afterglow phase. Especially, some of events will be correlated with
observed flares. These signals may be expected from not only long GRBs
but also short GRBs, that also may accompany flares \cite{Bur1}, if
such flares are
baryonic origins. However, several authors recently discussed flares
may not be of baryonic origins but of magnetic origins \cite{Zha3}. If
the outflow is much magnetized, formed shocks would be greatly weaken
and neutrino emission would be suppressed \cite{Zha3}. The 
detection of high energy neutrinos is one of the tests for the origin
of flares. In addition, these signals would give us information not only
on the magnetic field but also on the photon field. As demonstrated in
FUV-ray Flare (B) in Fig. 2, too copious photon field will reduce the
maximum proton energy and the following maximum neutrino
energy. Moreover, such neutrino detection may include
signals from FUV-ray flares, that are not seen due to 
absorption by neutral hydrogen both in host galaxy and in our Galaxy. 
Combined with gamma-ray large area space telescope (GLAST) mission
which may detect sub-GeV flashes by IC 
\cite{Fan1} or $\pi^{0}$ decay, these high energy neutrinos
may be important as a probe of FUV-ray emissions.

We have estimated the flux of neutrinos from flares by normalizing the
proton flux with the typical prompt radiated energy. For the capability of 
detections, outflows may need to be largely baryon loaded. 
Many flares are likely to be interpreted as the late activity of the
central engine. Such an energy injection is also one of the common 
interpretations for early flattening of the x-ray afterglow and its
energy will be comparable to that of the prompt burst
\cite{Zha2}. These late activities might supply the extra nonthermal
baryons. However, too large nonthermal baryon loading will not be
plausible \cite{KM1} and would also be constrained by high energy
$\gamma$ rays. The high energy $\gamma$ rays could cascade in the
source and/or in infrared and microwave background where the delayed
emission would occur \cite{Der1}. The delayed emission, if it occurs, would be
expected in the GeV-TeV region and extended to the keV-MeV
region. In addition, there would also be contributions of synchrotron
radiation components from charged particles such as pions.
Although the detailed calculation is needed to obtain reliable
spectra, such emission could not be detected
unless the source is very nearby and/or the flare is energetic. [For
example, in the case of the set with $L_{\mr{max}}={10}^{49}$ ergs/s,
$r={10}^{14.9}$ cm, and $\Gamma=30$, the inferred spectra \cite{Der1}
imply that the detection by burst alert telescope would be difficult
at $z \arangle 0.1$.] GLAST
would be able to test the existence of high energy emission.

At present, we do
not know the total radiated energy from flares very well.
The fluences of x-ray flares are usually smaller than that of the
bursts, but can be even comparable to or higher than that of the
bursts. There are also some GRBs with multiple flares, and there 
might be a significant fraction of the FUV-ray flares. We also do not know the
opening angle of flaring outflows themselves. If the late slow
outflow might have the larger opening angle than the prompt fast
outflow, of which beaming factor
$f_{b}=E_{\mr{GRB}}/E_{\gamma,\mr{tot}}
^{\mr{iso}}$ is typically $\sim (0.01-0.1)$, we can expect high energy 
neutrinos uncorrelated with GRBs.
Many parameters are still uncertain but these distributions are
important for more realistic predictions. For larger collision radii
with $r \sim {10}^{16}$ cm, we expect fewer neutrinos. Therefore, our 
evaluation could be maybe applied to about only a half of flares, and
the other half would make GeV flares and very high energy cosmic
rays. Unknown features of flares will be unveiled through more
multiwave-length observations by Swift, GLAST and so
on. We expect future neutrino observations will also give us some
clues on the physical parameters of flares.

Waxman and Bahcall \cite{Wax1} predicted neutrino burst under the
assumption that GRBs are the main sources of ultra-high-energy cosmic 
rays (UHECRs). In the case of the prompt bursts, the optical thickness for the
photomeson production can be smaller than the unity especially at
larger radii $r \arangle {10}^{14}{(E_{\gamma,\mr{sh}}^{\mr{iso}}/{10}^{51} \,
\mr{ergs})}^{1/2} \, \mr{cm}$ and the UHECRs can be
produced in such regions. In flares, it is more difficult to 
generate UHECRs especially in the case of FUV-ray flares where the
optical thickness for the photomeson production can be larger. 
So far, we have not taken account of neutrino oscillations. Actually, 
neutrinos will be almost equally distributed among
flavors as a result of vacuum neutrino oscillations \cite{Wax1}.

We thank the referees, K. Ioka, K. Toma, K. Asano, and Z.G. Dai for 
many profitable suggestions. S.N. 
is partially supported by Grants-in-Aid for Scientific Research from 
the Ministry of E.C.S.S.T. of Japan through No. 16740134.}
%\appendix

%\newpage %Just because of unusual number of tables stacked at end
%\bibliography{apssamp}% Produces the bibliography via BibTeX.

\begin{thebibliography}{}
\bibitem{Wax1}E. Waxman and J. Bahcall, Phys. Rev. Lett., 78, 2292
(1997); Phys. Rev. D, 59, 023002 (1998).
\bibitem{KM1}K. Murase and S. Nagataki, Phys. Rev. D, 73, 063002
(2006).
\bibitem{Wax2}E. Waxman and J. Bahcall, ApJ, 541, 707 (2000); 
Z.G. Dai and T. Lu, ApJ, 551, 249 (2001); C.D. Dermer, ApJ, 574, 65
(2002); C.D. Dermer
and A. Atoyan, Phys. Rev. Lett., 91, 071102
(2003); D. Guetta, D.W. Hooper, J. Alvarez-Mu\~niz, F. Halzen,
and E. Reuveni, Astropart. Phys., 20, 429 (2004); K. Asano, ApJ,
623. 967 (2005). 
\bibitem{Zha1}B. Zhang and P. M\'esz\'aros, IJMP. A, 19, 2385 (2004); 
T. Piran, Rev. Mod. Phys., 76, 1143 (2005).
\bibitem{Zha2}S. Kobayashi et al., astro-ph/0506157; 
B. Zhang \textit{et al}., ApJ, 642, 354 (2006); P.T. O'Brien
\textit{et al}., astro-ph/0601125.
\bibitem{Bur1}D.N. Burrows \textit{et al}., Sci., 309, 1833 (2005); 
S. Barthelmy \textit{et al}., Nat., 438, 994 (2005); G. Cusumano 
\textit{et al}.,
Nat., 440, 164 (2006); A.D. Falcone \textit{et al}., ApJ, 641, 1010 (2006); 
P. Romano \textit{et al}., A\&A, 450, 59 (2006).
\bibitem{Fan1}Y.Z. Fan and D.M. Wei, MNRAS, 364, L42 (2006);
Y.Z. Fan and T. Piran, MNRAS, 370, L24 (2006).
\bibitem{Ago1}S. Agostinelli \textit{et al}., Nucl. Instrum. Methods
Phys. Res., Sect. A, 506, 250 (2003).
\bibitem{PDG1}S. Schadmand, Eur. Phys. J. A, 18, 405 (2003); 
Particle Data Group, http://pdg.lbl.gov/.
\bibitem{Ahr1}ANTARES Collaboration, astro-ph/9907432 (1999);
P.K.F. Grieder \textit{et al}., Nuovo Cimento, 24C, 771 (2001);
J. Ahrens \textit{et al}., Astropart. Phys., 20, 507 (2004).
\bibitem{Iok1}K. Ioka, S. Kobayashi, and B. Zhang, ApJ, 631, 429 (2005).
\bibitem{Pro1}D. Proga and M. Begelman, ApJ, 592, 767 (2003); A. King
\textit{et al}., ApJ, 630, L113 (2005); R. Perna, P.J. Armitage, and B. Zhang,
ApJ, 636, L29 (2006).
\bibitem{Li1}K. Asano and F. Takahara, PASJ, 55, 433 (2003); 
Z. Li and L.M. Song, ApJ, 608, L17 (2004).
\bibitem{Jon1}F.C. Jones, Phys. Rev., 137, B1306 (1965).
\bibitem{Por1}C. Porciani and P. Madau, ApJ, 548, 522 (2001);
D. Guetta, T. Piran, and E. Waxman, ApJ, 619, 412 (2005).
\bibitem{Zha3}B. Zhang and S. Kobayashi, ApJ, 628, 315 (2005), 
Y.Z. Fan \textit{et al}., ApJ, 635, L129 (2005); Z.G. Dai et al.,
Sci., 311, 1127 (2006); D. Proga and B. Zhang, MNRAS, 370, L61 (2006).
\bibitem{Der1}C.D. Dermer and A. Atoyan, A\&A, 418, L5 (2004); 
S. Razzaque, P. M\'esz\'aros, and B. Zhang, ApJ, 613, 1072 (2004).
\end{thebibliography}
\end{document}